\documentstyle[epsf,epsfig,aps,prl,twocolumn]{revtex}
\begin{document}
\title{
Production Mechanism for Quark Gluon Plasma in Heavy Ion Collisions
}
\author{Ambar Jain and V. Ravishankar}
\vskip 0.5cm
\address{Department of Physics, IIT Kanpur, Kanpur-208 016, INDIA
}
\maketitle
\begin{abstract}
A general scheme is proposed here to describe the production of semi hard and soft
quarks and gluons that
form the bulk of the plasma in ultra relativistic heavy ion collisions.
We show how to obtain the production rates in the extended phase space - including the colour part -
as a function of time in a consistent manner, and without having to make ad-hoc assumptions.
All the required
features - the back reaction on QCD vacuum, the non-Markovian nature of the production,
and the quasi particle nature
of the partons are naturally incorporated.
We illustrate the
results with a realistic albeit toy model and also show how physically tenable
source terms may be obtained.
\end{abstract}

\maketitle
\narrowtext
There is a general consensus that the Quark Gluon Plasma (QGP) phase
of the Hadronic matter has already been
produced in ultra relativistic heavy ion collisions (URHIC) at CERN.
Future measurements
at RHIC \cite{rhic} and LHC \cite{lhc} are expected to provide more information on QGP.
Be it as may, it is still premature to claim that we have a full grasp of the
space-time history of QGP.
Indeed, a considerable effort has gone in  addressing the issue of evolution of QGP; starting with
the pioneering work of Bjorken \cite{bj} which was itself based on a highly prescient  work of
Landau \cite{landau},
 a number of results have been reported \cite{first,list,report}.
All these approaches are based on the
idea that the bulk of the plasma forms a classical gas, and evolves via an
appropriate transport equation.
Robust that this approach is \cite{lee,latt}, it has to be admitted that our
understanding of the production
process is not satisfactory. Recall that the transport equations (set up by incorporating
The Yang-Mills (YM) dynamics)
require  source terms that
need to be specified in the extended phase space (including the
compact part corresponding to the colour degree) as a function of time;
their derivation consequently
requires care: the basic
uncertainty principle should not be violated, and the multiple overlapping time scales that
pervade the phenomenon should not
be ignored. They should correctly reflect the dynamical nature of the vacuum \cite{bloch},
the non-Markovian nature
of the production rates \cite{schmidt}, and the quasi-particle nature
of the excitations \cite{weldon}.
It is needless to say that a source term lacking in one or more of the above
of the requirements
would predict incorrect
and unreliable features for the plasma.

Lattice computations \cite{nonp} suggest that even at   twice the transition temperature
$T_C$, the quarks and gluons do not
behave as ideal gases. We thus need effective field approaches based
on non-perturbative QCD. The
transport equation approach mentioned above provides a means of introducing collective degrees of
freedom appropriate for describing the non-perturbative dynamics, via the notion of a  background
chromoelectric field (CEF) which acts  as a source for the production of the bulk of QGP. The
acceleration of the partons in this field takes into account the long range interactions amongst
them. The short range part of the interaction can then be suitably modelled by an appropriate
collision term. That these aspects cannot be captured by the pQCD description \cite{pqcd} has been
emphasized recently \cite{shuryak}. We thus pay attention to the source terms obtained from the CEF
degrees of freedom.

The pioneering
work in Ref. \cite{bloch,schmidt,recent} introduces the non-Markovian nature into the
production process,
 by  employing  the Schwinger mechanism \cite{schwinger} for particle production.
This mechanism is valid only for static electric field configurations. Production
of particles from a space time dependent field would be described perturbatively
by the Schwinger Dyson expansion \cite{IZ};
it is well known from transport studies \cite{first,list,tdsource} that
the  CEF in URHIC is necessarily and {\it strongly}
space-time dependent by virtue of energy momentum conservations. See \cite{tdsource} for
a graphic comparison of the relative contributions of the Born and the Schwinger terms.
A remedy would be to correctly invoke an appropriate
perturbative mechanism \cite{tdsource} (over the nonperturbative vacuum with CEF),
 but unfortunately this study ignores the
non-Markovian aspect and also makes unjustified
assumptions in order to extract the time dependence of the source term.
Both the above attempts completely ignore the
essential dynamical nature of the colour charge, which is central to the non-abelian plasma
that we are interested in.
There has been some attempt \cite{list} to enlarge the phase space to include the colour part;
this attempt again involves
the  Schwinger mechanism apart from a lack of rigour in the derivation of the
colour dependence.
Finally, none
of the above (and other) approaches acknowledges the quasi particle nature of the excitations.
In short, the source terms obtained so far have utmost limited utility, and
a simple framework containing all the required features is still lacking.

 In this paper, we outline  a general  scheme for deriving
such a  source term, based on
simple but general considerations. (i) The non-perturbative aspects of the mechanism are
modelled after the colour flux tube model \cite{hagedorn}
 which is
known to provide a natural setting for discussing quark confinement\cite{thooft}. (ii)
The rate is extracted  not
 from an S-matrix computation, but by studying the time dependent evolution
 of the state in the Fock space.
As a natural byproduct, the quasi particle  and the
non-Markovian nature emerge automatically.

We now describe the flux tube model  in brief.
Soon after the two nuclei collide, and  start receding from each other, colour
strings are formed between them. These
strings merge to form the so called colour rope, a picturesque way of
stating that a  CEF is
 produced between the nuclei.
The consequent production process tantamounts to  the instability
of the QCD vacuum in the
presence of a classical
 CEF which is, in general, space time dependent.
Of course, the dependence itself is to be
obtained self consistently from a transport equation \cite{first,list};
here we merely content ourselves by indicating how to
infer the source term given an explicit history of CEF.
We study both the $q\bar{q}$ and the gluon productions in the presence of a classical
 YM field below. We discuss
The gluon case mainly since
its production process is quintessentially non-abelian with no counterpart in QED.

\noindent{\bf Formalism:}
Consider the gluon case first.
The gauge potentials $A_{\mu}^a$ are expanded as a sum of the classical values $C_{\mu}^a$
and their fluctuations
$\phi_{\mu}^a: <A_{\mu}^a> = C_{\mu}^a$. Expanding the YM Lagrangian,
the terms that are responsible for the
gluon production are identified to be
\begin{eqnarray}
L_{2g} &= &-\frac{g}{2} f^{abc}\big [(\partial_{\mu}C_{\nu}^a -  \partial_{\nu}C_{\mu}^a)
            \phi^{\mu b}  \phi^{\nu c} \nonumber \\
       & &  +       (\partial_{\mu}\phi_{\nu}^a -  \partial_{\nu}\phi_{\mu}^a) (C^{\mu b} \phi^{\nu c}
        +     \phi^{\mu b} C^{\nu c}) \big ]+{\cal O}(g^2)
\end{eqnarray}
where we have kept terms quadratic in the fluctuations.
The production rate is to be determined from the above
Lagrangian, by a suitable choice of the background field $C_{\mu}^a$.

Unlike in the Maxwell case, the YM configuration is not
completely specified by the field tensor \cite{wu}.
It is well known \cite{brown} that the consequent vacuum polarization
also differs for each of the
above (physically distinct) configurations. It is thus pertinent to examine which are the
configurations that would make the vacuum
unstable against particle production.
We have checked that {\it a la} the results for constant CEF \cite{brown}
only the `abelian configuration'
of the CEF leads to the quark and gluon productions - atleast in the leading order.
Keeping in mind the model and invoking (local) gauge covariance of CEF, we take the
 form of the gauge potential to be
  $C_{\mu}^a = \delta_{\mu, 0} \sum_iC_i(t, \vec{r})\delta_{a,i} $
where the summation is restricted  to the diagonal generators of the gauge group.
For all such configurations,
the magnetic field vanishes.
 Denote as usual the initial vacuum state, at $t=0$, by $\vert 0>$.
 Let $ \vert \Psi >(t) \equiv U(t,0) \vert 0>$ be the state at any later time $t$:
$U(t,0)$ is the evolution operator determined by the standard Schwinger-Dyson
expansion that is obtained from Eqn.1. Adapting the analysis in \cite{IZ},
the amplitude for the gluon production as a function of time is obtained by projecting
$ \vert \Psi > (t) $ to
the two gluon state   $\vert gg> \equiv \vert \vec{p}_1, \vec{p}_2; s_1,s_2; c_1,c_2 >$
labelled by momentum,
spin and colour quantum numbers
respectively. In the leading order which we consider for simplicity\cite{gen}, the amplitude is given by
\begin{eqnarray}
<gg \vert T(t) \vert 0 > = \frac{ig}{(2 \pi)^3} \frac{(E_2 -E_1)}{2 \sqrt{E_1 E_2}}
                          \vec{\epsilon}^{s_1}(\vec{p}_1) \cdot \vec{\epsilon}^{s_2}(\vec{p}_2)f^{a c_1 c_2}\nonumber \\
                         \tilde{C}^{0,a}(E_1 +E_2; \vec{p}_1 + \vec{p}_2; t)
\end{eqnarray}
where $T(t) \equiv U(t,0) -1$. Further,
$$
\tilde{C}_0^a =
              \int^t_0 dt_1 e^{-i(E_1 +E_2)t_1} \int d^3 \vec{r} exp (i(\vec{p}_1 + \vec{p}_2)
              \cdot \vec{r})
              C^{0,a}(t_1, \vec{r})
$$
is the incomplete Fourier transform of the gauge field and
$E_i$ are the energies carried by the gluons.
The lower limit of integration emphasizes that the CEF gets created at $t=0$ .
The corresponding expression for the $q\bar{q}$ production is given by
\begin{eqnarray}
<q\bar{q}\vert T(t) \vert 0> = \frac{g}{(2\pi)^3} \frac{m}{\sqrt{E_1E_2}}
                 \tilde{C}_0^a T^a_{c_1, c_2} u^{\dagger}_{s_1} ({p}_1)v_{-s_2}(-\vec{p}_2),
\end{eqnarray}
with  $\vert q\bar{q} > \equiv \vert \vec{p}_1, \vec{p}_2; s_1,s_2; c_1,c_2 >$.
$T^a$ are the generators of the gauge group in the fundamental representation,
while $u, v$ are the usual Dirac
spinors. For $t$ finite,
the gluons and the quarks that are in the final state
have non-overlapping projection with the unphysical degrees of freedom.
We  distinguish the physical pairs by projecting the states further onto the physical subspace,
by imposing the on the mass shell and the transversality restrictions as  subsidiary conditions
(analogous to the separation of the transverse part
of the electromagnetic field from the
longitudinal part in the study
of radiation by charged particles).

The probability that a pair is produced {\it any time} during the interval $ (0,t)$ is given by
$\vert <\phi \vert T(t) \vert 0 > \vert^2$, $\phi$ standing for either $gg$ or $q\bar{q}$.
 The production rate at any time $t$ is thus given
by its derivative at that instant;
clearly, the rate takes both positive and negative values, from which one concludes that the
particles acquire a (time dependent) finite life time.
Equivalently, vacuum has a dynamical role - both as a source and a sink. Further,
the rate at any time depends not merely on the field
configuration at that instant - it has a highly
non-Markovian character, being determined by the entire history.
In short, quantum interference effects dominate
the production mechanism, and this cannot be accessed by the approaches based on
Schwinger mechanism and pQCD.

\noindent{\bf A model example:}
We now illustrate the main results with an example drawn from its behaviour expected in
real situations \cite{bj}. Taking the gauge group to be $SU(3)$, let the CEF have the form
${\cal E}^a_i = \delta_{i,z}{\cal E}_0(\delta_{a,3}+ \delta_{a,8}) exp\{(\vert z \vert -t ) /t_0 \} \theta (t) \theta (t^2-z^2)$
with the initial field strength given by
${\cal E}_0$. This choice captures most of the features expected of a
boost invariant description \cite{bj}, although
it is by itself not a boost invariant configuration.
Notice that the field is characterised by a single time scale $t_0$,
in the units of which all the observables will be expressed. Indeed, $t_0$ gives
the decay time of the field or equivalently, the overall `production time' of the partons,
while $g$ is related to
the string tension.  The  initial energy density $\cal{U}$,
$t_0$ and $g$  are not entirely independent of each other:
$g ~{\cal U}~ t_0^4 \sim 1$.
We shall discuss the gluonic case in detail.
Consider the spin summed rate in the momentum space (see below for the colour part description).
We display the two particle rates because
eventually they are the ones needed to study signals involving correlations such as
dilepton production.
Since the rate is
defined in a six dimensional space, two dimensional sections will be
plotted as filled contour diagrams
after scaling them with respect to appropriate powers of $t_0$, $E_0^2$ and  $g^2$.
As an example, we fix
the total centre of mass energy $\sqrt{s}=0.6$ and  the
transverse momentum $p_T=0.1$. The rates are then exhibited
as contour plots in the plane defined by time and the longitudinal momentum $p_{L1}$
of one of the particles.
The longitudinal momentum $p_{L2}$ of the other particle then has two
solutions, one with the same sign as $p_{L1}$, and the other having the opposite sign.
We designate  the two channels by (+,+) and (+,-) respectively, and show in
figs. 1-4 the rates in both the
channels.
 Observe the following features:
The rate is a highly fluctuating function of all the variables involved,
changing signs rapidly, both with  the momenta as well as time, as shown in Figs 1-4.
The rate oscillates over a much larger range for smaller values of $p_T$ at a given $\sqrt{s}$.
For example, it oscillates roughly between
$ (-500,500)$ at $\sqrt{s}=0.6,~p_T=0.1$ (Figs. 1, 2). In contrast,
the range is restricted to
$\sim (-100,100)$
 when $p_T$ is increased to $0.2$ (Figs. 3, 4). Notice also that for a given $\sqrt{s}$ and $p_T$,
the fluctuations are more dominant for very small and very
large values of longitudinal momenta.
We have checked that the rate and its fluctuations are larger
at small values of $\sqrt{s}$, dropping rapidly as $\sqrt{s}$ is increased.
Though not shown here, the number
of oscillations are also seen to  increase as we increase $\sqrt{s}$.
It may also be seen from the figures that the (+,+) channel is  relatively
placid in comparison with the (+,-) channel.
However, the activity  at intermediate values of longitudinal momenta is more pronounced in the
(+,-) channel for a given value of $\sqrt{s}, p_T$.
In short, these figures exhibit in a vivid manner
the  production and the the absorption of the pairs,
and the correlation between them, in the momentum space.
The correlations are of vital importance in getting the screening lengths as well as in studying
the hadronization. Coming to the quark production, suffice it to mention that they are equally
rich in features differing only in the topology of the distributions. We do not show them here.

The analysis is incomplete in that the rates are still not described in the phase space.
This has been a vexed issue, with solutions attempted by imposing additional conditions
rather arbitrarily \cite{bj,casher,first}.
We resolve this issue by projecting the amplitudes to a basis of coherent states in the
extended phase space, a method that allows as accurate  a classical description as possible
for these quantum processes. As an illustration, consider the colour part which is unique
and central
to the plasma under discussion.
The structure of Eqns.(2,3) guarantees that it is independent
of momentum and spatial variables.
Let the gauge group be $SU(2)$ for simplicity (it has a simple topology).
Denoting the two gluon colour state in the coherent basis by
$\vert \hat{n}_1, \hat{n}_2 >$ \cite{perel},
 the colour part of the source term may be written as
 $\Sigma^g(\hat{n}_1, \hat{n}_2) \sim \vert \hat{n}_1 \times \hat{n}_2 \vert^2$
This expression has been obtained by averaging
over the direction of the CEF in the group space. Contrast $\Sigma^{g}$ with its counterpart for
$q\bar{q}$ production which has the structure
$\Sigma^q(\hat{n}_1, \hat{n}_2) \sim \frac{1}{2}( 1- \hat{n}_1 \cdot \hat{n}_2)$. While the gluons
are predominantly produced in directions normal to each other in the colour space, the quark pair
prefers to be produced in antiparallel directions, a feature that holds for $SU(3)$ as well, although
in a more involved fashion due to its more complicated topology.
Coupled with the fact that the colour charges
are indeed vectors in the group space and that they are dynamical variables obeying the
Wong equation \cite{wong}, we conclude that the dyanmics of QGP would differ significantly in the
bosonic and fermionic sectors in a manner that cannot be captured by simply multiplying the Maxwell
results by the number of degrees of freedom.
Finally, we have
performed a preliminary
computation and found that the phase space dependences of quarks and gluons in position and
momenta  bear no
resemblence to each other;
Those details will be presented  in a separate communication.


\begin{references}
\bibitem{rhic} J. L. Nagle Pramana, {\bf 37}, 355 (2001).
\bibitem{lhc} see, e.g., K. Werner, H. J. Drescher, S. Ostapchenko and
              T. Pierog, Nucl. Phys. {\bf A698}, 387 2002
\bibitem{shuryak} see, e.g., C. M. Hung and Edward V Shuryak, Phys. Rev. {\bf C57 }, 1891 (1998);
                             Diettrich Bodekar, Nucl. Phys.  {\bf B 559},502 (1999).
  \bibitem{bj} J. Bjorken, Phys. Rev D {\bf 27}, 140 (1983).
\bibitem{landau} L. D Landau, Akad. Nauk. SSSR. Ser. Fiz, {\bf 17}, 51 (1953);
                 E. Fermi, Prog. Theo. Phys., {\bf 5} 570 (1951);
                 I. Ya. Pomeranchuk, Dokl. Akad. Nauk. SSSR, {\bf 78}, 889 (1951).
 \bibitem{first} B. Banerjee, R. S. Bhalerao and V. Ravishankar, Phys. Lett., {\bf B224} 16 (1989) and
                references  therein.
\bibitem{list} see Gouranga C Nayak and V. Ravishankar,  Phys. Rev. {\bf D55}, 6877 (1997);
                 Phys. Rev. {\bf C58}, 356 (1998);
\bibitem{report}D. F. Litim and C. Manuel, Phys. Rept. {\bf 364} 451 (2002) and references therein.
 \bibitem{lee} P. Carruthers and F Zacharaiasen, Rev. Mod. Phys. {\bf 55}, 245 (1983);
               T. D. Lee, unpublished CU-TP-94;
                Dmitri Kharzeev, preprint hep-ph 0204014 (BNL-NT-02-05);
              P. F. Kelly, Q. Liu, C. luchessi and C. Manuel, Phys. Rev. Lett., {\bf 72}, 3461 (1994);
                Melissa Lampert and Benjamin Svetitsky, Phys. Rev. {\bf D61}, 0304011 (2000).
\bibitem{latt} There are proposals to simulate the space time evolution and
                determine the transport coefficients in
                URHIC on lattice: see, e.g., Sunao Sakai, Atsushi nakumara and Takuya Saito,
                Nucl. Phys. Proc. Suppl. {\bf 106}, 543 (2002);
                W. P\"{o}schl, Nucl. Phys. {\bf A661}, 641 (1999);
                S. A. Bass, B. M\"{u}ller and W. P\"{o}schl, J. Phys. G {\bf L109}, (1999).                   .                .
\bibitem{bloch} J. C. R. Bloch {\it et al}, Phys Rev. {\bf D60}, 116011 (1999)
\bibitem{schmidt}S. Schmidt {\it et. al.},Phys. Rev. {\bf D59}, 0940005 (1999);
                 J. C. R. Bloch, C. D. Roberts and S. M. Schmidt,
                  Phys. Rev. {\bf D61}, 117502 (2000).
\bibitem{weldon} Peter A Henning and E. Quack, Phys. Rev. Lett., {\bf 75}, 3811 (1995);
                  H. Arthur Weldon, unpublished (Hep-Ph 9809330);
                  I. V. Andreev, Mod. Phys. Lett.A {\bf 14}, 459 (1999)
\bibitem{nonp} G. Boyd {\it et al}, Phys. Rev. Lett. {\bf 75}, 4169 (1995);
               Nucl. Phys. {\bf B469}, 419 (1996)
\bibitem{recent} R. Alkhofer {\it et. al.}, Phys. Rev. Lett., {\bf 87}, 193902 (2001);
                 D. V. Vinnik {\it et. al.}, Euro. Phys. Jour. {\bf C22}, 341 (2001)
\bibitem{tdsource} R. S. Bhalerao and V. Ravishankar, Phys. Lett., {\bf 409}, 38 (1997).
\bibitem{schwinger} J. Schwinger, Phys. Rev. {\bf 82}, 664 (1951).
\bibitem{IZ} Claude Itzyksohn and J B Zuber, Quantum Field Theory, Mcgraw Hill (Singapore),
             191 -193 (1985).
\bibitem{pqcd} Klaus Geiger, Phys. Rev. {\bf D47},133 (1993).
\bibitem{hagedorn} T. S. Biro, H. B. Nielsen and J. Knoll, Nucl. Phys. {\bf B245}, 449 (1984).
                   For a recent review see, B. Svetitsky, eprint hep-ph/9907278
\bibitem{thooft} G. 't Hooft, Nucl. Phys. {\bf B190}, 455 (1981)
\bibitem{scully} see {\it e.g.}, Marlon O Scully, and M. Suhail Zubairi, Quantum Optics,
                 Cambridge University Press (Cambridge, UK) 1997
\bibitem{exp} The method employed here may therefore be of applicability to atomic systems.
               For a similarly interesting
              example, see  C. D. Roberts {\it et. al.},
                 Phys. Rev. Lett. {\bf 89}, 153901 (2002)
\bibitem{wu} T. T. Wu and C. N. Yang, Phys. Rev. {\bf D12} 3843 (1975)
\bibitem{brown} Lowell. S. Brown and William I Weisberger, Nucl. Phys. {\bf 157}, 285 (1979);
                 erratum: ibid. {\bf B172}, 54 (1980).
\bibitem{gen}  The expansion
              parameter here is $g {\cal E}_0$ where ${\cal E}_0$ is the initial field strength.
\bibitem{casher} A. Casher, H. Neuberger and S. Nussinov, Phys. Rev., {\bf D20},179 (1979)
\bibitem{perel} A. Perelomov, Generalized Coherent States and Their Applications, Springer-Verlag
             (Berlin, Germany) 1986
\bibitem{wong} S.K. Wong, Nuovo. Cim. {\bf A65}, 689 (1970).


\end{references}
\end{document}